\documentstyle[seceq,epsf]{ptptex}

\title{
Photon Absorption on a Neutron-Proton pair in $^3$He and $^4$He
}

\author{
Yukiko {\sc Umemoto}, Satoru {\sc Hirenzaki}
and Kenji {\sc Kume}
}

\inst{
Department of Physics, Nara Women's University, Nara 630-8506, Japan
\\
}

\recdate{
\today
}

\abst{
We have developed a phenomenological quasi-deuteron model to 
describe photodisintegration reactions of $^3$He and $^4$He 
at intermediate photon energies, and compared the results with 
experimental data obtained by TAGX group in which 
contributions from photon absorption by $pn$ pair were selectively
observed.  The data can be reproduced reasonably well, however 
there still remain certain discrepancies between calculated 
results and data.
}

\begin{document}

\maketitle

\section{Introduction}

Photodisintegration reactions of light nuclei have been studied
for a long time both theoretically and experimentally.
Experimental results of these reactions are naively
expected to be reproduced precisely by theoretical
works since the reactions are induced by the well-known
electromagnetic interaction and the structure of few nucleon
system is believed to be calculable.  Despite of this expectation,
there still remain certain discrepancies between experimental
data and theoretical results. We are interested in the
origins of these discrepancies.

The photodisintegration of nucleus was studied theoretically
using the quasi-deuteron model by Levinger \cite{Levinger51}
and later by Futami and Miyazima \cite{Futami71}.
They expressed the photodisintegration cross section of
nucleus using those of deuteron with a certain factor, 
so called Levinger's factor, which
accounts essentially for two effects, the relative
wavefunction of $pn$ pair and the effective
number of deuteron in the target nucleus.
Recently, the Levinger's factor was updated and determined
for many nuclei using the modern data of rms radius
\cite{Tavares92}, which enables us to obtain information of the 
relative wavefunction of $pn$ pair. 
The accuracy of the model was also investigated
for wide range of momentum transfer for trinucleon system
\cite{Gangopadhyay92}.  These results reproduced the experimental 
data qualitatively well indicating that the photon
absorption by $pn$ pair is important in photonuclear reactions.

On the other hand, the deuteron photodisintegration reaction,
which is treated as the elementary process in the quasi-deuteron
model mentioned above, was studied in microscopic way by
Leidermann and Arenh\"ovel \cite{Leidemann87}.
They calculated the two-body photo break-up of the deuteron
above pion threshold including $\Delta$ degrees of freedom
explicitly.  Homma and Tezuka also developed another
microscopic model of the deuteron photodisintegration
reaction \cite{Homma86}.  These models described
experimental data well at intermediate energy region,
E$_{\gamma} \ge 150 \sim $200MeV.

The microscopic models
were applied to ($\gamma$, p) reaction for several nuclear
targets assuming the quasi-deuteron picture
\cite{Homma86,Tezuka88}.
In the applications, they assumed two nucleon photon absorption 
processes expressed microscopically, and obtained the spectra 
of ejected protons
after absorbing the photon, including the distortion
effects as the attenuation factor.
They did not include the contributions from neither
spectator protons nor ejected protons after multiple scatterings 
which are expected to contribute in lower momentum region 
of proton spectra.
Thus, their models could only be applied to emitted protons with 
higher momenta.

Since 1987, new experimental data of the photodisintegration
reactions of the light nuclei have been obtained by TAGX group
\cite{Maruyama96}.  They have obtained kinematically complete
data of $^3$He($\gamma$,pn)p and $^4$He($\gamma$,pn)d for 
the first time\cite{Emura94,Maruyama97}.
They have striven to separate the whole events into two-
and multi- nucleon absorption processes by the momentum
ordering method \cite{Endo93}.  We are very interested in the
data since the contributions only from the photon absorption
by $pn$ pair are selectively observed which should be calculated
precisely in the quasi-deuteron picture, and
since the data are completely exclusive which do not
include any extra particles like pions in the final state.
Thus, we think it extremely important to investigate
this data to know whether it can be understood theoretically
or not.

So far, the $^3$He($\gamma$,pn)p reaction was studied
theoretically by Wilhelm $et$ $al$. \cite{Wilhelm94,Niskanen95}.
They considered both triplet (i.e. deuteron channel) and 
singlet configurations for
initial $pn$ pair in the nucleus, and included photon interaction
with nucleonic, mesic, and $\Delta$ currents. They calculated
the cross sections and spin observables, and found that the data
of total cross sections deviated from the theoretical results
at lower energies (E$_{\gamma} \le$ 200MeV).
Since their results for differential cross sections were not in
the form which could be compared with the data directly,
there have been no comparisons between data and theory for angular
dependent observables yet \cite{Niskanen95}.

The $^4$He($\gamma$,pn)d data \cite{Maruyama97} show that the
total cross section has strong energy dependence around
E$_{\gamma}$ = 150MeV which have not been investigated
theoretically. Theoretical results by Tezuka \cite{Tezuka88}
are consistent to the data for higher energies
but there are no theoretical calculation which can be compared
to the data in whole incident photon energy region.  Their 
results include the attenuation factor to account
for the distortion effects and the final
nuclear state is not specified to deuteron.

In this paper we attempt to understand these
reactions, $^3$He($\gamma$,pn)p$_{spectator}$ and \\
$^4$He($\gamma$,pn)d$_{spectator}$,  theoretically. 
For this purpose we investigate both $^3$He and $^4$He 
photodisintegraion reactions within a same theoretical framework
and try to
reproduce all data of the both reactions simultaneously.
We calculate all observables in suitable form which
can be compared to the data directly. Since
whole events are classfied carefully by
kinematical information of all final particles in the TAGX
experiments, we could also have
possibilities to check their classifcation scheme of the data
independently by studying the theoretical results assuming the
two body photoabsorption. The present work using the
phenomenological model is expected to give important guidelines
to fully microscopic studies which should be done as a next step.

We describe our model in the section 2.
Numerical results are compared with the data in section 3.
We summarize this paper in section 4.

\section{ Formalism }

In this section we describe our phenomenological model which is
applied to both $^3$He($\gamma$,pn)p$_{spectator}$ and
$^4$He($\gamma$,pn)d$_{spectator}$ reactions.
Since we would like to calculate all observables
in suitable forms to compare with the data directly,
we evaluate the final-state three-body phase space exactly.
The cross section for A($\gamma$,pn)B reaction can be
written as;

\begin{eqnarray}
d\sigma=\frac{1}{\lambda^{1/2}(s,0,M^2_A)}\mid T_A \mid ^2
        \frac{1}{(2\pi)^5}
        \delta^4(p_B+p_p+p_n-p_A-p_{\gamma}) \nonumber \\
\times \frac{M_B M_p M_n M_A}{E_B E_p E_n} d^3{\bf p}_B
d^3{\bf p}_p d^3{\bf p}_n,
\end{eqnarray}
where A indicates the target nucleus, $^3$He and/or $^4$He,
and B the spectator particle in the final-state, p$_{spectator}$
and/or d$_{spectator}$. The $s$ corresponds to the Mandelstam
variable in the initial $\gamma$+A system.  We integrate the
final-state phase space properly for each experimental result.

The $\lambda (\cdots)$ is the K\"allen function, which is
introduced to normalize the cross section properly by the
initial photon flux, defined as;

\begin{equation}
\lambda(a,b,c)=a^2+b^2+c^2-2ab-2bc-2ac .
\end{equation}

In our model, we consider only deuteron channel for the initial
$pn$ pair in the nucleus and express the square of the amplitude
of photon absorption by the nucleon pair using the differential 
cross section $\left(\frac{d\sigma}{d\Omega}\right)$ of 
d($\gamma$,p)n reaction as; 

\begin{equation}
\mid T(\gamma pn \rightarrow pn) \mid^2=
\left( \frac{d\sigma}{d\Omega} \right)
\lambda^{1/2}(s,0,M_d^2)(2 \pi)^2\frac{E_p+E_n}{M_p M_n M_d}
                \frac{1}{ \mid {\bf p}_p \mid},
\end{equation}
where all kinematical variables are evaluated in the
center of mass frame.  We use the phenomenological fit of 
the $\left(\frac{d\sigma}{d\Omega}\right) $ data \cite{Rossi89}, 
which reproduce the d($\gamma$, p)n cross section  well 
in the wide energy region, E$_{\gamma}$=20-440 MeV.

In order to calculate the observables for He target cases, we
postulate the quasi-deuteron picture and describe the square of
the $\gamma$ absorption amplitude by He nucleus as;

\[
\mid T_A \mid ^2=\mid T(\gamma pn \rightarrow pn) \mid^2 \mid
\psi({\bf p}_B)_{pn-B} \mid^2,
\]
where we evaluate the $\mid T(\gamma pn \rightarrow pn) \mid^2$
using the Mandelstam's $s$ and $t$ variables for two nucleon
system which absorb the photon.
The $\psi$ is the relative wave function
between $pn$ pair and spectator particle B
in He nucleus, which is assumed to be the Gaussian form as;

\begin{equation}
 \psi({\bf p}_B)_{pn-B}=
 ( 2\pi)^{3/2} \left( \frac{2}{p_f^2 \pi} \right) ^{3/4}
\exp \left( -\frac{p_B^2}{p_f^2} \right)   ,
\end{equation}
where $p_f$ means the Fermi momentum between $pn$ pair and
the spectator particle B and is roughly 
estimated to be 146 MeV for $^3$He
and 205 MeV for $^4$He from the data of the charge radii.
${\bf p}_B$ is evaluated in the laboratory frame. 

Here, we would like to describe some features of the present 
model. The advantages of our model are that
 we can calculate any cross sections which can be compared
to the data directly for both $^3$He and $^4$He target cases and 
that we can calculate observables in whole energy region of
interest using the experimental information of the deuteron
photodisintegration reactions.
However, we do not include neiter the effect of the deuteron 
spectroscopic factors nor the effects of the compact wave 
function of the $pn$ pair in the target He.
Both effects are expected to be insensitive to the incident 
photon energy and to change the overall normalization of the 
cross sections. 
Thus, we feel free to introduce a constant factor to each 
reaction as a parameter to normalize the absolute value of 
the cross sections when it is necessary.
Actually this factor is considered only in the case of
$^4$He($\gamma$,pn)d$_{spectator}$ reaction as we will see later. 
For $^3$He target cases, it is not necessary to consider this 
normalization factor and this could be accidental.
In the model, we do not include the contributions from the
singlet $pn$ pair in the initial state,
which were evaluated in refs. \cite{Wilhelm94,Niskanen95} 
and shown to be around 50 $\mu$b at E$_{\gamma}$ = 100 MeV 
for $^3$He($\gamma$,pn)p$_{spectator}$ total cross section. 
This contribution decreases monotonically as a function of 
E$_{\gamma}$ and is known to give minor contribusion to the 
photodisintegration reactions at higher energies. 
We should also mention here that the final state interaction
is partly included in the model from the beginning
since we have used the fit to the data of deuteron case.
The final state interaction effects for spectator particles 
are roughly considered in section 3.3. 
We compare the results with existing data 
and try to understand featuers of the data. 

\section{Numerical Results}
\subsection{$^3$He($\gamma$,pn)p$_{spectator}$ reaction}
In this section we compare our calculated results with 
experimental data of the
$^3$He($\gamma$,pn)p$_{spectator}$ reaction taken by TAGX group
\cite{Emura94,Endo93}. 

First we consider double differential cross sections (DDCS) for 
emitted protons, $\left(\frac{d^2\sigma}{dp d\Omega}\right) $. 
We expect to be able to fix the $p_f$ value defined in the last 
section by the widths of the DDCS.  
In Fig.1, theoretical results of DDCS are comapred with the data
at three different angles for each incident photon energy.
In all cases in Fig.1, experimental data show two distinct peaks
which correspond to spectra due to the spectator protons and 
energetic protons paticipated in photon absorption process. 
The peaks at higher momenta have stronger angular dependence 
than the spectator contributions and become small rapidly at
large angles.  Our calculated results with $p_f$=150MeV
reproduce all DDCS data reasonably well
without any adjustable parameters.  In order to see the 
sensitivity of the calculated sepctra to the $p_f$ value, 
we also show the results with $p_f$=200 MeV in Fig. 1. 
We see that the spectra are affected by the $p_f$ value as 
we expected. However, it seems difficult to determine the $p_f$ 
value precisely from these data.  Thus, we use the
$p_f$=150MeV, which is consistent to the charge radius of $^3$He, 
in the following calculations.

The differental cross sections for protons
are shown in Fig. 2 for four incident
photon energies. Except for the $E_\gamma$=325 MeV case, 
the data are reasonably well reproduced again without 
any parameters.  At $E_\gamma$=325 MeV, the observed differential 
cross section is small at forward angles in contrast to 
calculated results.  The differential cross sections
are insensitive to the $p_f$ value and are qualitatively the 
same for
all $p_f$ values which are consistent to the DDCS data shown 
in Fig. 1.

We show the total cross sections of the $^3$He 
photodisintegration by two body ($pn$ pair) photon absorption 
process in Fig. 3 as a function of the incident photon energy. 
Experimental total cross sections are obtained by integrating 
the differential cross section data in Fig. 2 over the observed 
kinematical region. The contribution from unobserved kinematical 
region is corrected for by extrapolation using the two-nucleon 
absorption model \cite{private}.  
Theoretical results only include the deuteron channel of initial 
$pn$ pair in the target. 
As we can see from the figure, the data show the peak at 
E$_{\gamma}$=225MeV, which seems due to the $\Delta$ excitation, 
while the calculated results at 275MeV. This difference 
could be explained by inclusion of the siglet configuration of 
$pn$ pair in the initial state which was evaluated by Wilhelm 
$et$ $al$. \cite{Wilhelm94,Niskanen95} and shown to be larger 
for lower photon energies.  However, the energy dependence of 
the theoretical results will become more different from the data 
by including the singlet contribution at $E_{\gamma}$=125 $\sim$ 
205 MeV. In this energy region, the data increase monotonically 
with photon energy while the theoretical results may be flat or 
even have opposite energy dependence by including the 
singlet contribution. 
Our results are consistent to those calculated by Wilhelm $et~al$
\cite{Wilhelm94,Niskanen95} for initial $pn$ pair in the deuteron 
channel.

\subsection{$^4$He($\gamma$,pn)d$_{spectator}$ reaction}
Total cross sections of the $^4$He($\gamma$,pn)d$_{spectator}$ 
reaction were measured at several photon energies by TAGX group 
\cite{Maruyama97} and shown to have strong energy dependence 
around $E_\gamma$=150MeV. This energy dependence is almost like 
a 'discontinuity' to lower energy data taken by different groups 
\cite{Doran,Arkatov,Gorbunov,Balestra}.  We first calculate the 
total corss section with $p_f$=200MeV, which is consistent to 
the charge radius of $^4$He.  We show the results in Fig. 4 and 
Fig. 5 together with the data. As can be seen in both figures, 
the calculated results are larger than the data at lower energies 
and smaller than the data at higher energies, and
do not reproduce the strong energy dependence at 150MeV.

Then, we vary the $p_f$ in wide range to investigate the 
possibility to reproduce the energy dependence by changing 
the wavefunction of $^4$He.  We also multiplied a factor to 
the calculated cross sections to reproduce the $\Delta$ peak. 
The results are shown in Fig. 6 
and Fig. 7.  We find that the our results do not reproduce the 
experimental energy dependence.  It is difficult to reproduce the 
observed 'discontinuity'.

We would like to mention here the recent work by
Efros $et$ $al$. \cite{Efros97}, in which the total $^4$He
photodisintegration cross section is studied in the giant 
resonance energy region. They have shown both the total and the 
two-body decay contributions, which are $^4$He($\gamma$,n)$^3$He 
and $^4$He($\gamma$,p)$^3$H reactions, separately for 
$^4$He photodisintegration 
cross section. We can estimate the absolute value 
of cross section of our case (3-body decay) by subtracting 
the 2-body decay contribution from the total.  It is about 
1 mb at E$_\gamma \sim$ 40 MeV which is even larger than our 
largest result($\sim$ 500 $\mu$b). 
If this is correct, the older data at lower energies may include 
large errors.

\subsection{Final state interaction effect}
In our quasi-deuteron model, since we have used 
the $d(\gamma,p)n$ data to 
evaluate the amplitude of the photon absorption process by 
$pn$ pair, the effects of the final state interaction between  
proton and neutron, which absorb the photon, is included. 
However, we have not included the distortion effects between 
the spectator and the other particles. 
We think it necessary to evaluate the final state interaction 
effects for spectator particles carefully for more realistic 
calculations. Since the two body photon 
absorption events are selected by kinematical information by 
TAGX group, it could be difficult to separate multi-nucleon 
photo-absorption events from two-body absorption events affected 
by the final 
state interaction, especially at low $E_\gamma$ cases. 
If the spectator has relatively larger momentum because of the
final state interaction, it could not be seen as the spectator.

In this paper, we have estimated the final state interaction 
effects in a simple way to see sensitivities of 
cross sections by changing the $\psi_{pn-B}$ defined 
in eq. (2.4) as follows;

\begin{equation}
 \psi^{\prime}({\bf p}_B)_{pn-B}=
\int d{\bf r}_B [1 - f \exp(-r^2_B/r^2_0)]
\exp ( -i {\bf r}_B \cdot {\bf p}_B ) \psi({\bf r}_B)_{pn-B}   ,
\end{equation}
where $\psi({\bf r}_B)_{pn-B}$ is the relative wavefunction 
between $pn$ pair and spectator particle B in coordinate space, 
of which momentum space expression is given by eq. (2.4). 
Parameters $f$ and $r_0$ are introduced to distort the plave wave 
of spectator particle, $\exp(-i {\bf r}_B \cdot {\bf p}_B)$, 
in the final state.  Results without 
distortion effects correspond to $f=0$. 

We calcluate total cross sections for both $^3$He and $^4$He 
cases with changing the parameters in the range of 
$-1 \le f\le 1$ and $0.5 \le r_0 \le 2.0$ [fm].  
We find that the absolute value of total cross sections changes
significantly by this modification. However, in the present 
simple model, energy dependence of the cross sections 
is almost the same as those with $f=0$ case. 
As a natural expectation, the distorion effects have 
energy dependence.
Thus, we need to include realistic energy dependence in the 
distortion effect for further studies. 

\section{Summary}
In this paper, we have investigated the
photodisintegration reactions $^3$He($\gamma$,pn)p$_{sp}$ and
$^4$He($\gamma$,pn)d$_{sp}$ which have been observed by TAGX 
group. 
We have investigated both reactions within a same theoretical 
model and comapre the calculated results with 
experimental data. In the model, we have assumed the 
quasi-deuteron mechanism to describe the photon absorption 
amplitude using the experimental information of deuteron 
photodisintegration. We have introduced the relative wave
function between $pn$ pair and spectator particle in the target 
He and treated the phase space integration carefully in order to 
calculate the observables in suitable coordinate system 
to compare with the data. 
Using the present theoretical model, we can calculate 
observables in whole energy region of interest and can compare 
with the experimenal data directly. 
The final state interacton is partly 
included in the model since we have used the fit to the data of 
deuteron photodisintegration.

Our model is found to reproduce gross features of all the 
existing data but certain descrepancies remain, which are
(i) energy dependence of the total cross section for 
$^3$He($\gamma$,pn)p$_{sp}$ 
and (ii) the step-like change of the total cross section 
of $^4$He($\gamma$,pn)d$_{sp}$ reaction around 
E$_\gamma\sim 150$ MeV.  We found that the these discrepancies 
are difficult to reproduce by the
present phenomenological model.  Thus, we think that these
discrepancies involve important information on the essential 
differences of the photodisintegration reactions of $^3$He and 
$^4$He from that of deuteron and should be treated carefully. 

For further studies, we need to develope a microscopic
model based on the accurate wave functions of initial and final states and
photon interaction with hadronic currents in order to understand 
these reactions deeper.  We think it is very important to 
calculate the observables in the suitable form to compare 
with data by the microscopic model.

\section*{Acknowledgments}
We would like to thank TAGX group for stimulating discussions.
Especially we would like to express our sincere thanks to 
Prof. K. Maruyama, Prof. T. Suda, Dr. S. Endo
and Dr. K. Niki for useful communications.

\vspace{1.8cm}
{\large Figure Caption}
\vspace{0.5cm}
\begin{quote}
Figure~1:
 Double differntial cross sections of the 
$^3$He($\gamma$,pn)p$_{sp}$ reactions for three differnt angles 
at four photon energies as indicated in figures. 
The solid and dashed lines indicate the calculated results 
with $p_f$=150MeV and $p_f$=200MeV, respectively. 
Data are measured by TAGX \cite{Endo93}. 
\end{quote}
\vspace{.5cm}
\begin{quote}
Figure~2:
 Differntial cross sections of the $^3$He($\gamma$,pn)p$_{sp}$ 
reactions at four photon energies as indicated in figuers. 
The solid lines indicate the calculated results with 
$p_f$=150MeV. 
Data are taken from ref. \cite{Endo93}.
\end{quote}
\vspace{.5cm}
\begin{quote}
Figure~3:
 Total cross section of the $^3$He($\gamma$,pn)p$_{sp}$ 
reaction is shown as a function of incident photon energy. 
The solid curve indicates the calculated result 
with $p_f$=150MeV. 
Data are taken from ref. \cite{Emura94}. 
\end{quote}
\vspace{.5cm}
\begin{quote}
Figure~4:
 Total cross section of the $^4$He($\gamma$,pn)d$_{sp}$ 
reaction is shown as a function of incident photon energy. 
The solid curve indicates the result with $p_f$=200MeV. 
Data are those from refs. \cite{Maruyama97}(solid circles), 
\cite{Doran}(squares), \cite{Arkatov}(crosses), 
\cite{Gorbunov}(open circles), \cite{Balestra}(triangles). 
\end{quote}
\vspace{.5cm}
\begin{quote}
Figure~5:
 Ratio of $\sigma(^4He(\gamma,pn)d_{sp})/\sigma(d(\gamma,p)n)$ 
as a function of E$_{\gamma}$. \\
$\sigma(^4He(\gamma,pn)d_{sp})$ is calculated with $p_f$=200MeV. 
Data are those from refs. \cite{Maruyama97}(solid circles), 
\cite{Doran}(squares), \cite{Arkatov}(crosses), 
\cite{Gorbunov}(open circles). 
\end{quote}
\vspace{.5cm}
\begin{quote}
Figure~6:
 Same as in Fig. 4. Each line indicates calculated result with 
$p_f$=100(dashed curve), 200(solid curve), 300(dotted curve), 
and 400(dash-dotted curve)MeV, respectively. 
Calculated results are normalized to reproduce experimental peak 
height at $\Delta$ energy region. These normalization factors are 
1.45 for $p_f$=100MeV, 1.53 for $p_f$ 200MeV, 
1.67 for $p_f$=300MeV, and 1.88 for $p_f$=400MeV. 
\end{quote}
\vspace{.5cm}
\begin{quote}
Figure~7:
Same as in Fig. 5. Each line corresponds to calculated results 
with different $p_f$ values and normalization factors as given 
in Fig. 6. 
\end{quote}














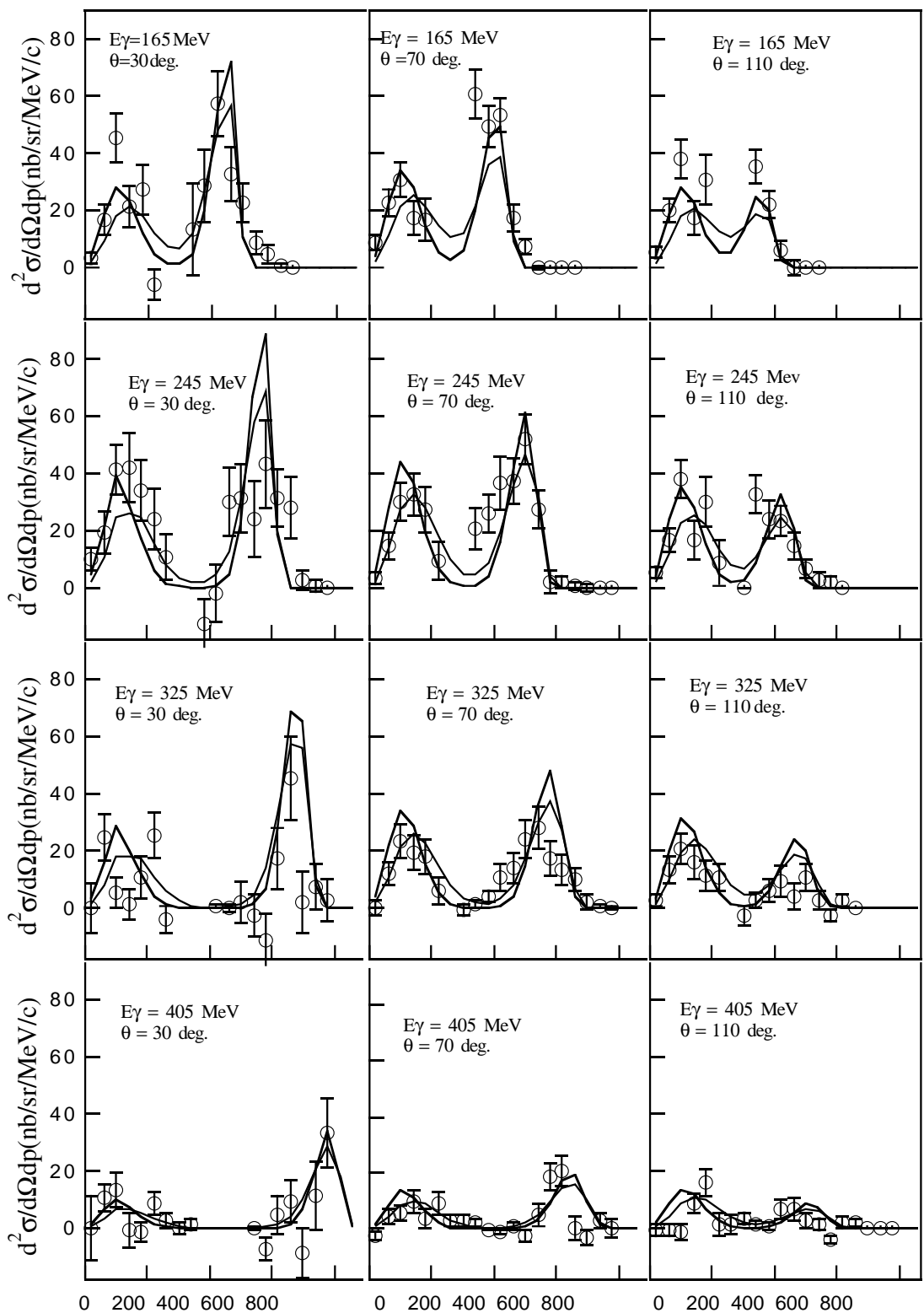
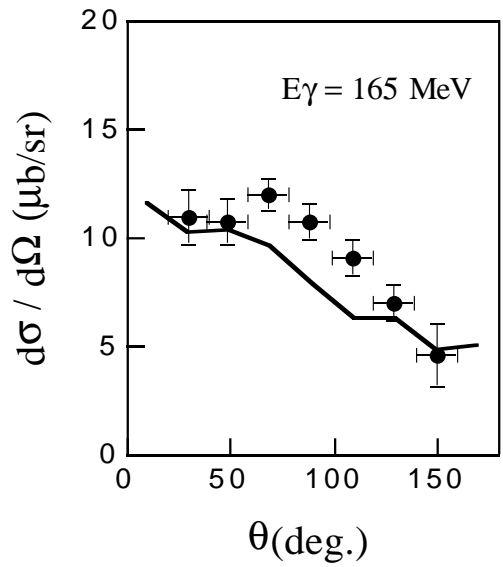
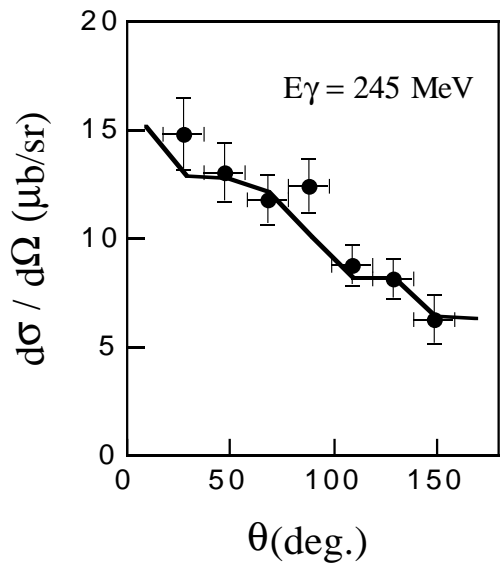
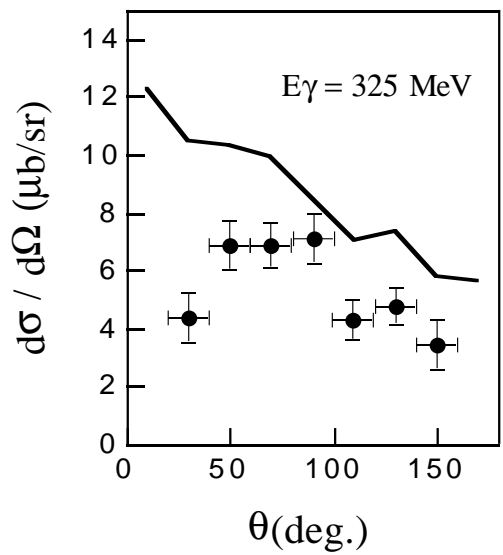
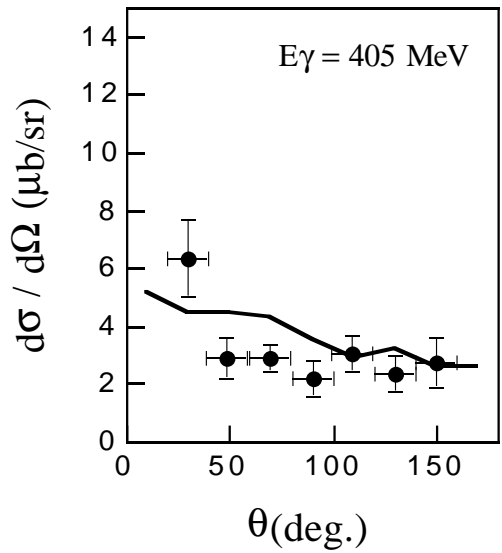
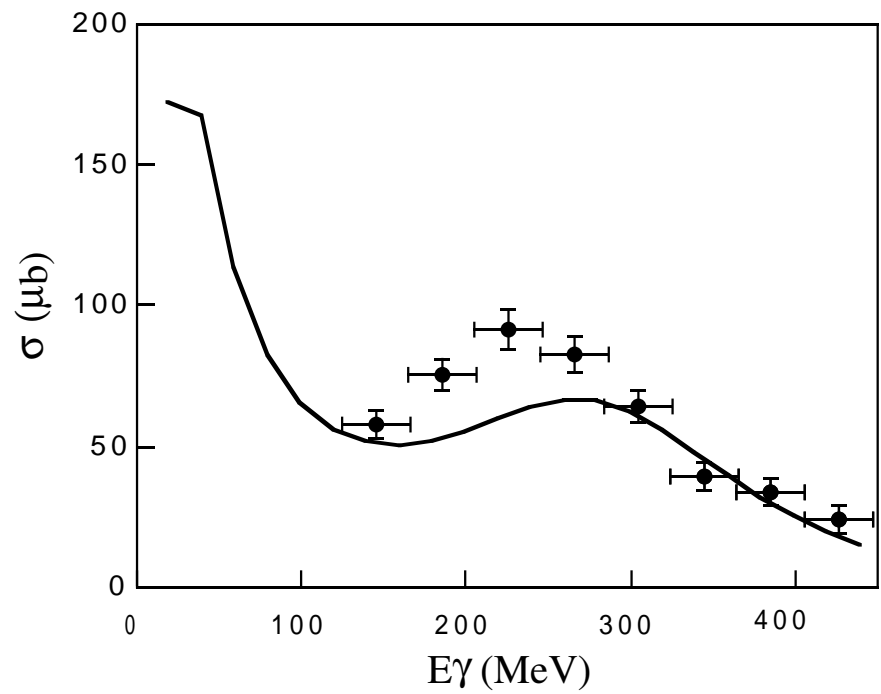
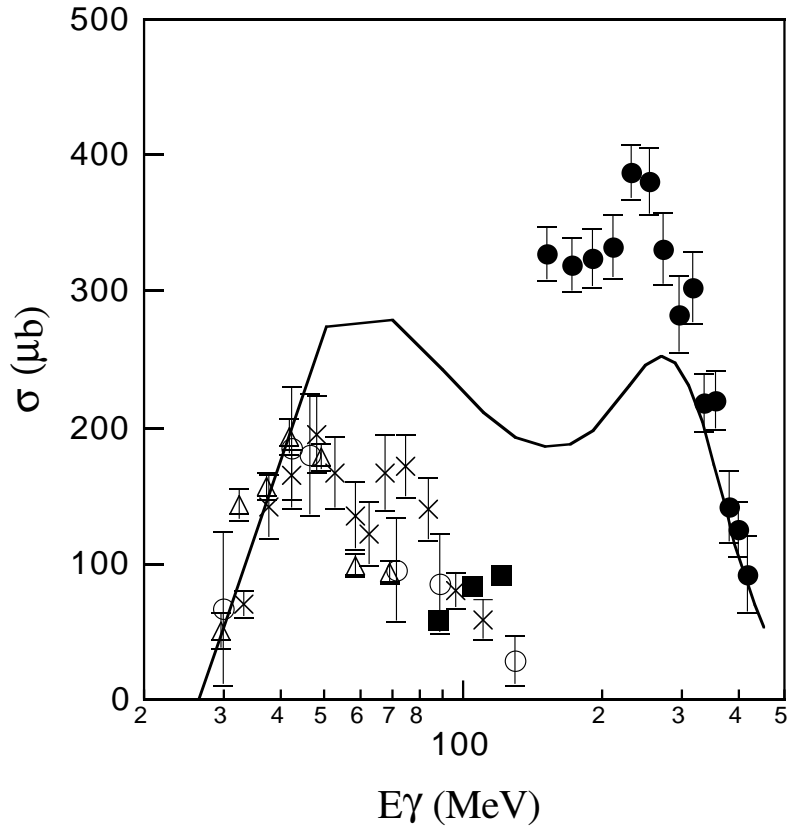
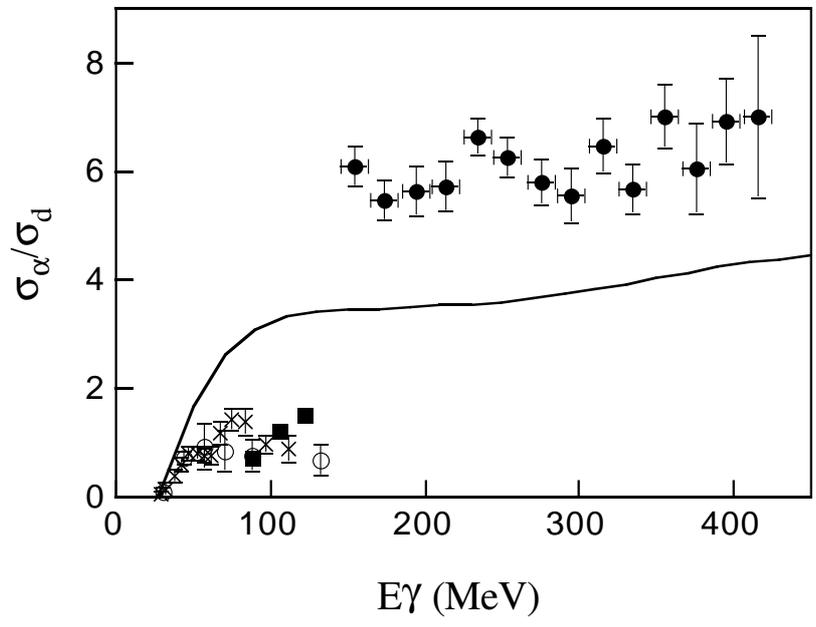
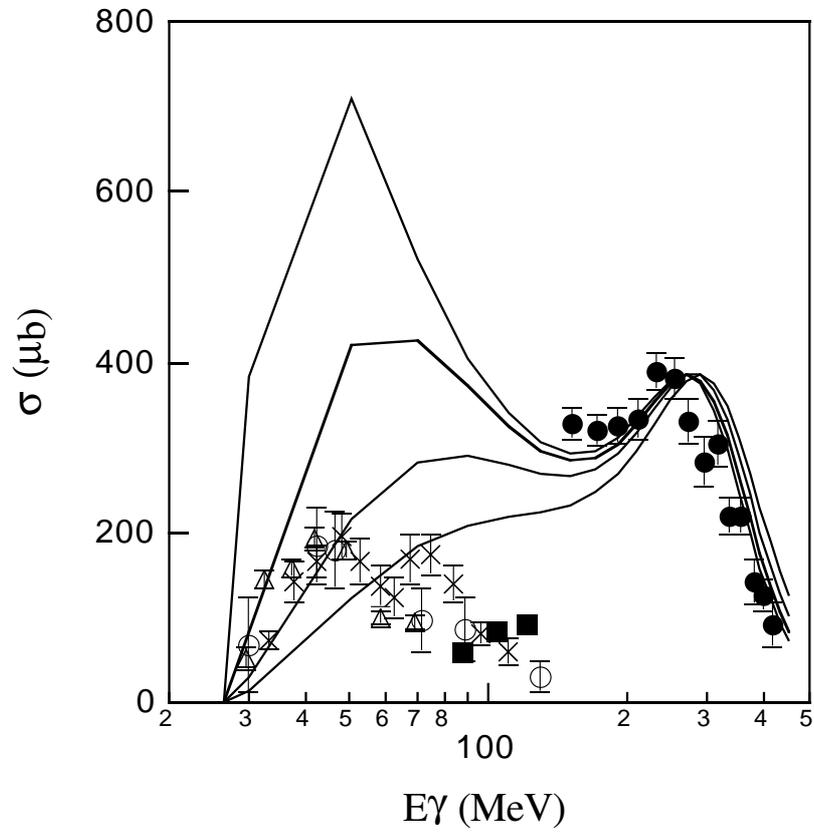
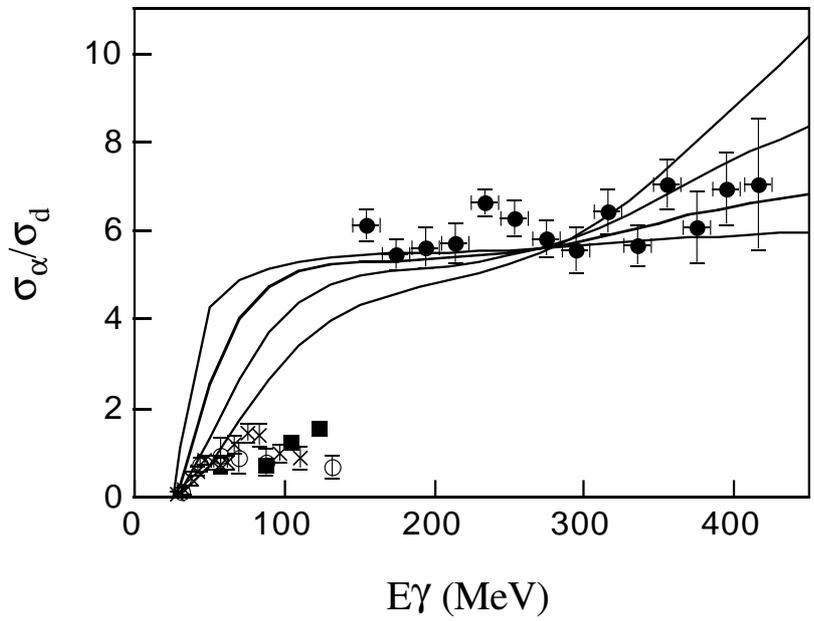
\end{document}